\documentclass{article}
\usepackage{spconf,amsmath,graphicx}
\usepackage{booktabs, subfig}
\usepackage{tablefootnote}
\usepackage{setspace}

\title{EXPLORING WAV2VEC 2.0 FINE TUNING FOR IMPROVED SPEECH EMOTION RECOGNITION}
\name{Li-Wei Chen, Alexander Rudnicky}
\address{Language Technologies Institute, Carnegie Mellon University}
\begin{document}
%
\maketitle
\begin{abstract}
While Wav2Vec 2.0 has been proposed for speech recognition (ASR), it can also be used for speech emotion recognition (SER); its performance can be significantly improved using different fine-tuning strategies.
Two baseline methods, vanilla fine-tuning (V-FT) and task adaptive pretraining (TAPT) are first presented.
We show that V-FT is able to outperform state-of-the-art models on the IEMOCAP dataset.
TAPT, an existing NLP fine-tuning strategy, further improves the performance on SER.
We also introduce a novel fine-tuning method termed P-TAPT, which modifies the TAPT objective to learn contextualized emotion representations.
Experiments show that P-TAPT performs better than TAPT, especially under low-resource settings.
Compared to prior works in this literature, our top-line system achieved a 7.4\% absolute improvement in unweighted accuracy (UA) over the state-of-the-art performance on IEMOCAP.
Our code is publicly available.\footnote{https://github.com/b04901014/FT-w2v2-ser}

\end{abstract}
\begin{keywords}
Speech emotion recognition, deep neural networks, wav2vec 2.0, fine-tuning, pretrained models
\end{keywords}
\section{Introduction}
\label{sec:intro}
Speech emotion recognition (SER) remains one of the key components in human-machine interaction and in human communication systems.
With the development of deep learning, several attempts~\cite{fayek2017evaluating,han2014speech,8659587} have been made to automatically learn emotion representations from audio signals using neural nets.
However, the improvement of deep learning based systems is often limited by the lack of annotated data.
Commonly used SER datasets~\cite{busso2008iemocap,SAVEE,Livingstone2018TheRA} are relatively small in size in comparison to automatic speech recognition (ASR) datasets.
Moreover, systems trained on these datasets may not generalize well to other domains such as call centers.

Self-supervised pretrained models~\cite{Devlin2019BERTPO,Lan2020ALBERT} provide a solution by first learning from a large scale speech corpus without explicit labeling.
The knowledge learned from pretraining can be transferred to downstream tasks by either using the model as a feature extractor or directly fine-tuning the whole model.
While first introduced for the purpose of natural language processing (NLP), several pretrained models~\cite{schneider19_interspeech,NEURIPS2020_92d1e1eb,hsu2021hubert} have been developed for speech processing. Wav2vec~\cite{schneider19_interspeech} is a multi-layer convolutional neural network (CNN) trained to predict future frames conditioned on past frames by minimizing a contrastive loss.
On the other hand, wav2vec 2.0~\cite{NEURIPS2020_92d1e1eb} is a transformer-based model that adopts a masked learning objective to predict missing frames from the remaining context.

Despite the success of these methods in ASR, speaker verification, and mispronunciation detection~\cite{NEURIPS2020_92d1e1eb,fan21_interspeech,peng21e_interspeech}, only a few attempts~\cite{Boigne2020RecognizingME,xia21b_interspeech,pepino21_interspeech} have been made to apply them on SER.
Boigne et al.~\cite{Boigne2020RecognizingME} find that wav2vec features are superior to traditional spectral-based features on SER.
Xia et al.~\cite{xia21b_interspeech} compare features extracted with different time spans and conclude that features with longer temporal context such as wav2vec perform better on SER.
Pepino et al.~\cite{pepino21_interspeech} show that features extracted from a linear combination of layers outperform singe layer representations in wav2vec 2.0 on SER.
While these studies demonstrated the usefulness of the pretrained models as feature extractors, little research has been conducted on fine-tuning them for SER.

One persistent issue on fine-tuning pretrained models is the mismatch between pretraining and target domain~\cite{dontstoppretraining2020,hsu21_interspeech}.
Task adaptive pretraining (TAPT)~\cite{dontstoppretraining2020} is proposed to resolve the domain shift by continuing the pretraining process on the target dataset.
Hsu et al.~\cite{hsu21_interspeech} show that TAPT greatly improves generalization and robustness on ASR when the pretraining and fine-tuning data are dissimilar.
Since the speech in the pretraining ASR corpus differs from emotive speech in multiple regards~\cite{Pell2001InfluenceOE}, we consider TAPT a compelling method for fine-tuning on SER.

\begin{figure*}
    \centering
    \includegraphics[width=\linewidth]{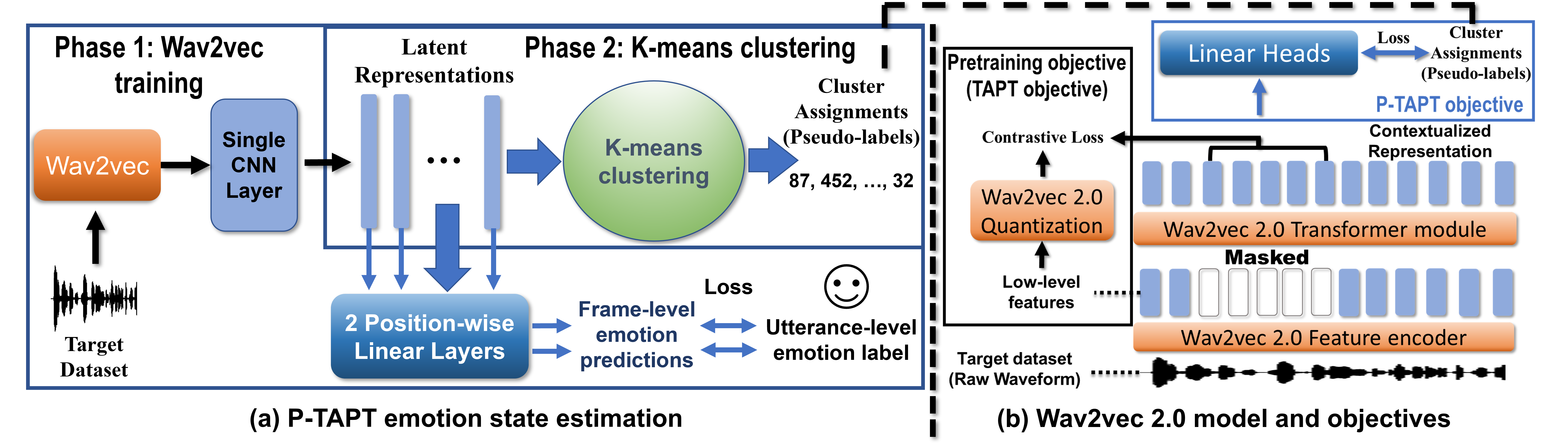}
    \caption{System overview of our methods. (a) Emotion state estimation phase of P-TAPT. An additional CNN with stride 2 is used to align the time steps between wav2vec and wav2vec 2.0. The output of cluster assignments will be used as pseudo-labels for the P-TAPT objective. (b) Model architecture and pretraining objective of wav2vec 2.0 along with our P-TAPT objective.}
    \label{fig:system}
\end{figure*}

In this paper, we explore methods for fine-tuning wav2vec 2.0 on SER.
We show that by adding a simple neural network on top of wav2vec 2.0, vanilla fine-tuning (V-FT) outperforms state-of-the-art (SOTA) methods on the IEMOCAP~\cite{busso2008iemocap} dataset.
In addition, with V-FT as a baseline, TAPT significantly boosts the performance of fine-tuning wav2vec 2.0 on SER.
Furthermore, motivated by previous works on the benefits of segment-based emotion features~\cite{fayek2017evaluating, DBLP:conf/interspeech/MaoCKL20, xia21b_interspeech} and self-supervised representation learning~\cite{hsu2021hubert,caron2018deep}, we developed a novel fine-tuning procedure for SER which yields even better performance, especially in low-resource conditions.
Finally, we achieve a 7.4\% absolute increase on unweighted accuracy (UA) over the SOTA performance on IEMOCAP.



\vspace{-1.2em}
\section{Method}
\label{sec:method}
We first review wav2vec 2.0, which serves as the backbone model for the methods we examine.
We then present the two baseline methods we established.
Finally, we introduce pseudo-label task adaptive pretraining (P-TAPT), a novel method we designed to fine-tune wav2vec 2.0 on SER.

\vspace{-0.9em}
\subsection{The wav2vec 2.0 model}
\label{ssec:w2v}
Wav2vec 2.0 is a transformer-based model trained to extract contextualized representations from raw audio signal.
Figure~\ref{fig:system}.b shows the wav2vec2.0 model architecture and its pretraining objective.
It consists of three sub-modules, feature encoder, transformer module, and quantization module.
Feature encoder is a multi-layer CNN that processes the input signal into low-level features.
Based on this representation, the transformer module is further applied to produce contextualized representation.
The quantization module discretizes the low-level features into a trainable codebook. 
To train the model, part of the low-level features are masked from the transformer module, and the objective is to identify the quantized version of the masked features based on its context.\footnote{There is an additional diversity loss in pretraining which promotes the diversity of the quantization codebook.}

\vspace{-0.9em}
\subsection{Comparing methods}
As there is no existing baseline system fine-tuning wav2vec 2.0 on SER, we created two baselines.
One is the conventional fine-tuning method, and the other is task adaptive pretraining which is first introduced in NLP.

\textbf{Vanilla fine-tuning.} Wav2vec 2.0 differs from its NLP counterparts~\cite{Devlin2019BERTPO} in that there is no utterance-level pretraining task to naturally form a sentence representation.
As a consequence, aggregation across time steps is required to fine-tune on utterance level classification tasks. 
We experimented with different configurations and found that using average pooling on the final layer is simple yet effective for SER.
Specifically, the final contextualized representation extracted by wav2vec 2.0 is first processed by a global average pooling across the time dimension, then followed by the ReLU activation and a single linear layer to predict the emotion categories.
In addition, a modified version of SpecAugment~\cite{Park2019} proposed in wav2vec 2.0 is applied during training for better generalization.
We will use this architecture for the fine-tuning stage of all three methods.
We abbreviate the vanilla fine-tuning method as V-FT.

\textbf{Task adaptive pretraining.} Task adaptive pretraining (TAPT)~\cite{dontstoppretraining2020} is a simple but effective method to fine-tune pretrained language models~\cite{Devlin2019BERTPO} on domain-specific tasks.
It bridges the difference between the pretraining and target domain by continuing to pretrain on the target dataset.
In this paper, we examine TAPT as one of the methods of fine-tuning wav2vec 2.0 on SER.
To distinguish from the original pretraining and fine-tuning stage, we define an intermediate task adaptation stage for the continual pretraining process.

\vspace{-0.9em}
\subsection{Pseudo-label task adaptive pretraining}
\label{ssec:pseudo}
While TAPT adapts to emotive speech by continual training with the pretraining objective, it does not make use of emotion labels.
Essentially, the contextualized representations obtained will be general features suitable for various downstream tasks.
As we only focus on SER, we propose to adapt this objective to generate emotion-specific features.
Instead of identifying the missing low-level features, we focus on predicting the emotion state of the masked sequence.
One advantage it brings is better data efficiency.
Reconstruction of missing audio parts is a more complicated task, which makes the model vulnerable to over-fitting.
Additionally, it simplifies the fine-tuning stage as it already filters out information unrelated to emotion recognition from the contextualized representation.

However, frame-level emotion states need to be recognized to realize our method.
While only utterance-level emotion labels are given for most of the SER dataset, several studies~\cite{xia21b_interspeech,fayek2017evaluating,DBLP:conf/interspeech/MaoCKL20} indicate that frame-level emotion information can still be inferred by training with a segment-based classification objective.
Particularly, as shown in Figure~\ref{fig:system}.a, we fine-tune wav2vec to extract frame-level emotion representation that is useful for predicting an utterance-level emotion label.
We find that using CNN architectures such as wav2vec is important since the locality of CNN preserves sequential structure.
After training, we run k-means clustering algorithm~\cite{1056489} on all of the extracted representations from the target dataset.
As Mathilde et al.~\cite{caron2018deep} have shown, the k-means cluster assignments on intermediate layers of CNN classifiers can capture information related to the target labels.
Therefore, we interpret this cluster assignment as a pseudo-label that represents the local emotion state.

We then replace the TAPT objective with our new P-TAPT objective.
We add a position-wise linear head composed of two linear layers to predict the k-means cluster assignments of the masked frames.
In practice, we run multiple k-means clustering with a different number of clusters, and our model needs to predict an ensemble of cluster assignments with multiple linear heads.
This cluster ensemble technique is shown to facilitate representation learning in HuBERT~\cite{hsu2021hubert}, a recent self-supervised speech representation learning model.

\vspace{-0.5em}
\section{Experimental Setup}
\label{sec:exp-setup}
\subsection{Dataset}
We use two datasets for evaluation, IEMOCAP~\cite{busso2008iemocap} and SAVEE~\cite{SAVEE}.
We only use the speech modality.

\textbf{IEMOCAP}.
Interactive Emotional Dyadic Motion Capture (IEMOCAP) is a popular dataset for evaluating SER systems.
It contains five recording sessions, each with one male speaker and one female speaker.
To compare with previous works, we use the default labels provided by IEMOCAP.
However, only four emotion categories are considered: neutral, sad, angry, and happy.
In particular, the ``excited'' category is merged with ``happy'' due to its sparsity in the dataset.
The total amount of speech is about 7 hours.

\textbf{SAVEE}.
The Surrey Audio-Visual Expressed Emotion (SAVEE) dataset contains four male speakers: DC, JE, JK, and KL.
Each speaker reads out the same set of 120 sentences labeled with one of the 7 emotion categories: angry, disgust, sad, fear, happy, surprise, and neutral.
We use all of the emotion categories, which results in 480 utterances with a total of 30 minutes of speech.

\subsection{Training and evaluation procedure}
All experiments use the same learning rate $1 \times 10^{-4}$ with Adam optimizer~\cite{kingma2014adam}.
For the wav2vec model, we use a pretrained model developed by Facebook AI\footnote{https://github.com/pytorch/fairseq/tree/master/examples/wav2vec}.
We build our wav2vec 2.0 implementation on top of the huggingface implementation and adopt a pretrained model checkpoint from Facebook AI\footnote{https://huggingface.co/facebook/wav2vec2-base}.
Both models are pretrained on the unsupervised speech of LibriSpeech 960h~\cite{7178964} without transcriptions.
We evaluate our systems using unweighted accuracy (UA)~\cite{han2014speech} under a speaker-independent setting; the speakers in the test set are excluded from the training data.
Additional implementation details are provided in our github repository.
We run each experiment 5 times for the full IEMOCAP dataset and 20 times for SAVEE and on sub-sampled versions of IEMOCAP.
Additionally, we observe that wav2vec 2.0 fails to converge with some of the random seeds.
Therefore we discard and rerun outlier runs where the performance is outside two standard deviations from the mean.

\textbf{IEMOCAP}. To have a fair comparison with the majority of previous works, we split the dataset by leaving one session out as test set, the remaining four sessions are used for training.
Note that most of the papers using IEMOCAP do not explicitly define their validation set~\cite{etienne18_smm}.
We therefore train with all four sessions for a fixed 15 epochs without validation using a batch size of 64.
The number of epochs is chosen so that each of our competing methods converges in terms of training loss.

\textbf{SAVEE}. A similar evaluation procedure is used for SAVEE.
In each fold, one speaker is left out for testing and the remaining three are used for training.
We increase the number of training epochs to 30 and the batch size is halved to 32 for the smaller training set.
\vspace{-0.5em}
\section{Results and Discussion}
\label{sec:result}

\subsection{Comparison of fine-tuning methods}
Table~\ref{tab:iemocap} compares performance for the fine-tuning methods on IEMOCAP.
For all sessions except the first, TAPT yields a noticeable improvement over V-FT, and P-TAPT performs better than TAPT for all sessions.
On the other hand, Table~\ref{tab:SAVEE} shows that on SAVEE, both TAPT and P-TAPT outperform V-FT by a large margin.
However, the performance of P-TAPT is very close to that of TAPT on SAVEE.
We analyze these results by considering the characteristics of SAVEE and IEMOCAP.

\textbf{Domain shift and linguistic content.}
We first quantify the domain shift between both datasets and the pretraining dataset.
We take the wav2vec 2.0 model pretrained on LibriSpeech and calculate the pretraining loss on both datasets along with the test set of LibriSpeech.\begin{table}[th]
  \centering
  \caption{Comparison of methods on IEMOCAP in UA(\%)}
  \label{tab:iemocap}
  \begin{tabular}{l c c c c c c}
    \toprule
    \multicolumn{1}{l}{\textbf{Session}} & \multicolumn{1}{c}{\textbf{1}} & \multicolumn{1}{c}{\textbf{2}} & \multicolumn{1}{c}{\textbf{3}} & \multicolumn{1}{c}{\textbf{4}} & \multicolumn{1}{c}{\textbf{5}} &  \multicolumn{1}{c}{\textbf{Mean}}\\
    \midrule
    V-FT & $71.0$ & $76.2$ & $66.3$ & $68.7$ & $67.3$ & $69.9$ \\
    TAPT & $71.8$ & $79.6$ & $70.2$ & $73.2$ & $72.5$ & $73.5$\\
    P-TAPT & $\textbf{72.8}$ & $\textbf{80.2}$ & $\textbf{71.0}$ & $\textbf{73.6}$ & $\textbf{73.7}$ & $\textbf{74.3}$\\
    \bottomrule
  \end{tabular}
\end{table}
\begin{table}[th]
  \centering
  \caption{Comparison of methods on SAVEE in UA(\%)}
  \label{tab:SAVEE}
  \begin{tabular}{l c c c c c c}
    \toprule
    \multicolumn{1}{l}{\textbf{Speaker}} & \multicolumn{1}{c}{\textbf{DC}} & \multicolumn{1}{c}{\textbf{JE}} & \multicolumn{1}{c}{\textbf{JK}} & \multicolumn{1}{c}{\textbf{KL}} &  \multicolumn{1}{c}{\textbf{Mean}}\\
    \midrule
    V-FT & $75.2$ & $78.8$ & $56.0$ & $39.0$ & $62.3$ \\
    TAPT & $81.6$ & $83.3$ & $\textbf{69.9}$ & $\textbf{49.7}$  & $\textbf{71.1}$\\
    P-TAPT & $\textbf{86.7}$ & $\textbf{84.2}$ & $66.8$ & $45.8$ & $70.9$\\
    \midrule
    Human & $73.7$ & $67.7$ & $71.2$ & $53.2$ & $66.5$\\
    \bottomrule
  \end{tabular}
\end{table}
Table~\ref{tab:pt-loss} verifies the presence of domain shift on both datasets providing room for TAPT to improve.
A smaller loss indicates that SAVEE is closer to LibriSpeech as the model can already generalize well to SAVEE.
However, this improvement is larger on SAVEE than IEMOCAP despite the smaller domain shift.
We observe a strong correlation between linguistic content and emotion labels in SAVEE.\footnote{Two-thirds of the sentences are specific to one emotion and shared across all speakers.}
We conjecture that this correlation is captured by our model and surpasses human evaluators who annotate emotion from only para-linguistic information.
This also explains why P-TAPT does not further improve TAPT, as the TAPT objective is already suitable for modeling linguistic information.
Nonetheless, in more naturally elicited emotional conversations (IEMOCAP), P-TAPT performs better than TAPT.

\textbf{Data efficiency.}
We also investigated the behavior of our methods when presented with different amounts of training data.
Specifically, We fix session 5 of IEMOCAP as the held-out test set, and gradually halve the number of training examples in the remaining four sessions by random selection.
We compare TAPT and P-TAPT using the ratio of their corresponding improvements over V-FT.
A lower ratio indicates that the improvement from P-TAPT is more significant than TAPT.
As shown in Table~\ref{tab:data-eff}, this ratio is lower under low-resource settings with one hour or less of training data.
Thus P-TAPT is more data-efficient than TAPT.
As mentioned in Section~\ref{ssec:pseudo}, we attribute this to the change of objective from the reconstruction of audio frames to the prediction of emotion states which is less data-intensive though it requires labeled data.

\begin{table}[th]
  \caption{Wav2vec 2.0 pretraining loss on different datasets}
  \label{tab:pt-loss}
  \centering
  \begin{tabular}{l c c c c c c}
    \toprule
    \multicolumn{1}{c}{\textbf{Dataset}} & \multicolumn{1}{c}{Libri.(test-clean)} & \multicolumn{1}{c}{SAVEE} & \multicolumn{1}{c}{IEMOCAP}\\
    \midrule
    \multicolumn{1}{c}{\textbf{Loss}} & $32.04$ & $41.37$ & $55.42$\\
    \bottomrule
  \end{tabular}
\end{table}
\vspace{-1.0em}
\subsection{Comparison with prior works}
Table~\ref{tab:comp} compares our performance on IEMOCAP to that of existing SOTA models.
We only include methods that evaluate under speaker-independent settings.
Simply fine-tuning the wav2vec 2.0 model (using V-FT) outperforms wav2vec 2.0 without fine-tuning\cite{pepino21_interspeech} by 3.6\% absolute UA.
The P-TAPT method provides 7.4\% absolute improvement over SOTA models on IEMOCAP.
We also show performance for methods that use both speech and
text~\cite{chen20b_interspeech,santoso21_interspeech}; our audio-only method appears comparable.
\vspace{-0.2em}
\begin{table}[th]
  \caption{Comparison of methods on data efficiency in UA(\%) on the session 5 of IEMOCAP}
  \label{tab:data-eff}
  \centering
  \begin{tabular}{l c c c c}
    \toprule
    Data size & $\sim$0.5hr & $\sim$1hr & $\sim$3hr & $\sim$7hr\\
    \midrule
    V-FT & $56.8$ & $60.9$ & $66.4$ & $67.3$\\
    TAPT & $58.1$ & $62.9$ & $69.3$ & $72.5$\\
    P-TAPT & $\textbf{58.8}$ & $\textbf{64.1}$ & $\textbf{70.0}$ & $\textbf{73.7}$\\
    \midrule
    Ratio(\%) & 65.0 & 62.5 & 80.6 & 81.3\\
    \bottomrule
  \end{tabular}
\end{table}
\vspace{-1.0em}
\begin{table}[th]
  \caption{Comparison with prior works on IEMOCAP}
  \label{tab:comp}
  \centering
  \begin{tabular}{l c c c c}
    \toprule
    \multicolumn{1}{c}{\textbf{Method}} & \multicolumn{1}{c}{\textbf{Feature}} & 
                                         \multicolumn{1}{c}{\textbf{UA} (\%)}\\
    \midrule
    FCN+Attention~\cite{8659587}  & Spectrogram & 63.9\\
    Wav2vec w/o. FT~\cite{Boigne2020RecognizingME} & Wav2vec\tablefootnote{\label{feat}We identify it as a feature instead of model architecture since they only use wav2vec/wav2vec 2.0 as feature extractor without fine-tuning.} & 64.3\\
    Wav2vec w. FT~\cite{xia21b_interspeech}  & Waveform & 66.9\\
    Wav2vec 2.0 w/o. FT~\cite{pepino21_interspeech} & Wav2vec 2.0 & 66.3 \\
    \midrule
    Wav2vec 2.0 w. V-FT & Waveform & 69.9\\
    Wav2vec 2.0 w. TAPT & Waveform & 73.5\\
    Wav2vec 2.0 w. P-TAPT & Waveform & \scalebox{1.1}{\textbf{74.3}}\\
    \midrule
    \midrule
    Audio + Text~\cite{chen20b_interspeech} & MFCC+ALBERT\tablefootnote{\label{bert_text}Referring to the text features extracted from ALBERT~\cite{Lan2020ALBERT} and BERT~\cite{Devlin2019BERTPO}.} & \textit{72.1}\\
    Audio + ASR~\cite{santoso21_interspeech} & MFCC+BERT & \textit{75.9} \\
    \bottomrule
  \end{tabular}
\end{table}
\vspace{-1.0em}
\section{Conclusion}
\label{sec:conclusion}
We describe different fine-tuning strategies for wav2vec 2.0 on SER. These strategies produce SOTA performance on IEMOCAP, a well-studied corpus.
We verify the presence of domain shift in SER and demonstrate that addressing it improves performance.
We describe an algorithm for learning contextualized emotion representation and show its advantage in fine-tuning a wav2vec 2.0 model for SER.
We believe that these techniques can be generalized to other tasks and can provide a basis for research on the utility of contextualized emotion representation.
We intend to continue exploring the usefulness of this approach, in a multi-modal setting.

\vspace{-0.8em}
\section{Acknowledgements}
We are grateful to PwC USA as well as to The Digital Transformation and Innovation Center at Carnegie Mellon University for supporting our research.
We thank Yangyang Xia and Richard M. Stern for the discussions and feedback.

\vfill\pagebreak

\let\OLDthebibliography\thebibliography
\renewcommand\thebibliography[1]{
  \OLDthebibliography{#1}
  \setlength{\parskip}{1.8pt}
  \setlength{\itemsep}{1.8pt}
}
\bibliographystyle{IEEEbib}
{\small\bibliography{main}}

\end{document}